\documentclass{iau}
\usepackage{graphicx}
\usepackage{float}
\usepackage{xcolor}

\newcommand{\msun}{M_\odot}
\newcommand{\nb}{\texttt{NBODY6++GPU}}
\newcommand{\rebound}{\texttt{REBOUND}}
\newcommand{\amuse}{\texttt{AMUSE}}

\newcommand{\lonelyplanets}{\texttt{LonelyPlanets}}

\title[IAUS345.~~ Dynamics of Planets in Star Clusters] 
{Planetary Systems in Star Clusters: the dynamical evolution and survival}

\author[Flammini, Cai, Spurzem \& Kouwenhoven]   
{F. Flammini Dotti$^{1,2}$, Maxwell Xu Cai$^3$,  \\ Rainer Spurzem$^{4,5,6}$ \and M.B.N. Kouwenhoven$^{1,2}$
}

\affiliation{$^1$Department of Mathematical Sciences, Xi{'}an Jiaotong-Liverpool University, 111 Ren{'}ai Rd., Suzhou Dushu Lake Science and Education Innovation District, Suzhou Industrial Park, Suzhou 215123, P.R. China \quad
  email: {\tt flammini.francesco@xjtlu.edu.cn} \\[\affilskip]
$^2$Department of Mathematical Sciences, University of Liverpool, Liverpool L69 3BX, UK \\
$^3$Leiden Observatory, Leiden University, PO Box 9513, 2300 RA, Leiden, Netherlands\\
$^4$National Astronomical Observatories and Key Laboratory of Computational Astrophysics, Chinese Academy of Sciences, 20A Datun Rd., Chaoyang District, 100012, Beijing, China\\
$^5$Kavli Institute for Astronomy and Astrophysics at Peking University, 5 Yiheyuan Rd., Haidian District, 100871, Beijing, China\\
$^6$Zentrum f\"{u}r Astronomie der Universit\"{a}t Heidelberg, Astronomisches Rechen-Institut, M\"{o}nchhofstr. 12-14, 69120 Heidelberg, Germany
}

\pubyear{2019}
\volume{345}  
\jname{Origins: from the Protosun to the First Steps of Life}
\editors{Bruce G. Elmegreen, L. Viktor T\'oth, Manuel G\"udel, eds.}
\begin{document}

\maketitle

\begin{abstract}
 Most stars, perhaps even all stars, form in crowded stellar environments. Such star forming regions typically dissolve within ten million years, while others remain bound as stellar groupings for hundreds of millions to billions of years, and then become the open clusters or globular clusters that are present in our Milky Way galaxy today. A large fraction of stars in the Galaxy hosts planetary companions. To understand the origin and dynamical evolution of such exoplanet systems, it is necessary to carefully study the effect of their environments. Here, we combine theoretical estimates with state-of-the-art numerical simulations of evolving planetary systems similar to our own solar system in different star cluster environments. We combine the \rebound{} planetary system evolution code, and the \nb{} star cluster evolution code, integrated in the \amuse{} multi-physics environment. With our study we can constrain the effect of external perturbations of different environments on the planets and debris structures of a wide variety of planetary systems, which may play a key role for the habitability of exoplanets in the Universe.

\end{abstract}

\begin{figure}[tb]
\centering
\begin{tabular}{cc}
  \includegraphics[width=0.4\textwidth,height=!]{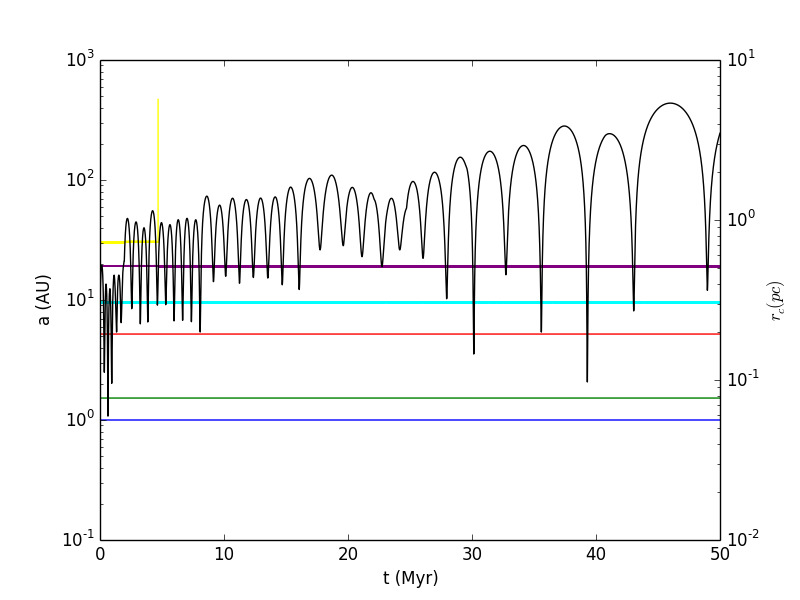} &
  \includegraphics[width=0.4\textwidth,height=!]{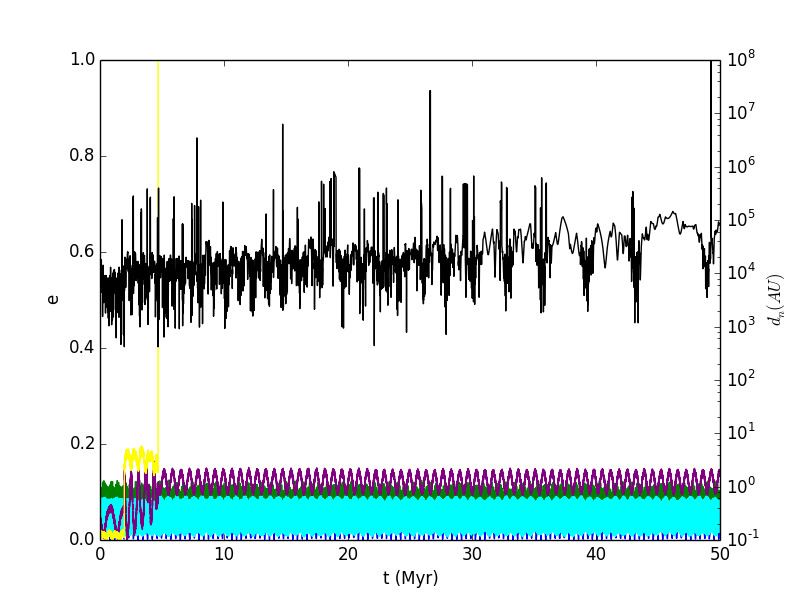} \\
  \end{tabular}
\caption{A typical example of a planetary system evolving under the influence of neighbouring stars (Flammini et al., in prep.). {\it Left:}  semi-major axis, $a$, and distance from the cluster centre, $r_c$. {\it Right:} eccentricity, $e$, and distance to the nearest neighbour star, $d_n$. The planetary systems were evolved using the \lonelyplanets{} code (\cite[Cai et al.2015, 2017; Flammini et al., in prep.]{cai2015,cai2017}), which combines the stellar dynamics (\nb; \cite{wang2016}) with that of the planetary systems (\rebound; \cite{rien2012}) in the \amuse{} framework (\cite{pelupessy2013}).  We model star clusters with $N=500-10^4$ stars and radii $r_{\rm vir}=1$\,pc in a Solar-orbit tidal field. Stars are drawn from the Kroupa\,(2001) IMF, with masses of $0.1-25\,\msun$. Stars in the mass range $0.9-1.1\,\msun$ are assigned a planetary system similar to our own Solar system ({\color{blue}Earth}, {\color{green}Mars}, {\color{red}Jupiter}, {\color{cyan}Saturn}, {\color{violet}Uranus} and {\color{yellow}Neptune}). All models are integrated for 50\,Myr.}
\label{figure:escapers}
\end{figure}


\noindent
A large fraction of stars in the Galaxy hosts planetary companions (e.g., \cite{mayo2018}). To understand the origin and evolution of planetary systems, it is necessary to study the influence of their surroundings. Most field stars form in dense gaseous environments, which are similar to, but slightly less massive and concentrated than the progenitor of the longer-lived open clusters. In such environments, planetary systems may be perturbed or disrupted following close encounters with neighbouring stars (e.g., \cite{spurzem2009, hao2013, zheng2015, cai2017}). Most of these young groups of stars disperse within $10-50$~Myr (e.g., \cite{lada2003}), after which their member stars become part of the field star population, while others remain bound for hundreds of millions to billions of years. Here, we present the results of our numerical study of the evolution of complex planetary systems (Solar-like Systems) in star clusters, in which we model the effects of nearby cluster member stars on existing planetary systems. We aim to study a different variety of surrounding environments, to obtain a comprehensive picture of the conditions under which external perturbations are important or negligible. 
The evolution of planetary systems is determined by (i) secular evolution due to planet-planet interactions, (ii) gravitational encounters with nearby stars, and (iii) the cumulative effect of distant encounters. The external contributions are primarily determined by the local stellar density: planetary merger rates and escape rates increase strongly with density. Most systems remain stable for at least 50\,Myr, although a substantial fraction experiences orbital reconfigurations, planet-planet mergers, and escape events. As a results of its high mass, Jupiter often serves as a barrier that separates dynamics in the inner and outer planetary system, and hence to some degree protects the terrestrial planets from external perturbations. In systems where Jupiter itself is strongly perturbed, however, the terrestrial planets are unlikely to remain in the habitable zone. Amongst the planetary systems that remain intact, we do not find differences between those that remain in the cluster, and those that have escaped into the Galactic field. On the other hand, we  find a notable correlation between the evolution of planetary systems and the velocity with which a star escapes from the cluster: planetary systems amongst ejected stars often experienced highly destructive events, while evaporated stars remain mostly intact. The fraction of ejected planets is comparable for all models, while the fraction of planet-planet mergers tends to increase with the local stellar density in the star cluster. For further details we refer to Flammini et al. (in prep.).

\vspace{-0.3cm}
\ \\
\noindent
{\em MBNK acknowledges the National Natural Science Foundation of China (grant 11573004) and the XJTLU Research Development Fund (RDF-16-01-16). RS and MXC acknowledge the support of the DFG priority program SPP 1992 "Exploring the Diversity of Extrasolar Planets" (SP 345/20-1), and of Silk Road Project at NAOC Beijing.}

\end{document}